\documentclass[a4paper,11pt]{article}
\usepackage{jheppub} 
\usepackage{lineno}

\title{\boldmath Prospects for Measuring the Higgs Boson Decay to WW$^\ast$ in Fully Hadronic Final States at the ILC using Multivariate Techniques}

\author{M. Pandurovi\'c}
\affiliation{Vin\v{c}a Institute of Nuclear Sciences,\\
Mike Petrovi\'ca Alasa 12-14, Belgrade, Serbia}

\emailAdd{milap@vinca.rs}

\abstract{In this paper, the statistical potential of the measurement of Higgs to WW$^\ast$ decay at the International Linear Collider (ILC) is presented. The Higgs boson of 125  mass is produced through the Higgsstrahlung production channel. The study is conducted at two center of mass energies, 250 and 500~\text{GeV}. The fully hadronic final state is analyzed. The analysis is performed on Monte Carlo data samples obtained using detailed ILD detector simulation, assuming integrated luminosity of 500 fb$^{-1}$ and maximal beam polarization of both beams,  $P_{e^+e^-}=(+0.3,-0.8)$. The background from $\gamma\gamma$ to hadron processes is overlaid over each event. Analyses are performed using machine learning. The obtained relative statistical uncertainties of the measurement are 4.1$\%$ and 6.5$\%$, at 250 and 500~\text{GeV}, respectively. }

\begin{document}
\maketitle
\flushbottom

\section{Introduction}
\label{sec:intro}

After the discovery of the Higgs boson, all measurements obtained at the LHC are confirmed to agree with the Standard Model predictions. However, there is a reasonable belief that the Standard Model is not a complete description of nature, suggesting that additional fundamental interactions are certain to be discovered. Among several methods to search for these new interactions, the search for deviations in the properties of elementary particles from those expected within the Standard Model is the most promising. In particular, the measurement of the Higgs boson couplings to the Standard Model particles is assuredly the most prominent candidate, since the experimental uncertainties of the analyses performed at the LHC leave room for the presence of new physics. Several beyond Standard Model theories foresee the deviations of the Higgs boson couplings. These theories provide the scale at which new bosons must appear and set the necessary precision goal, which is to be achieved, to ensure the required visibility \cite{Wells}. For most models, the upper limits of the expected couplings deviations are of the order at the level of 5$\%$ or less, varying as $m^{2}_{H}/M^2$ where M is the predicted mass of the new particles and M$_H$ is the mass of the Higgs boson \cite{ILDPhysicsCase}. 

The International Linear Collider \cite{ILC} is developed as a high-luminosity linear electron-positron collider based on the superconducting RF accelerating technology. At the first stage, with a 250~\text{GeV} centre of mass energy, the collider is designed to operate as the Higgs boson factory. The assumed instantaneous luminosity is 1.35$\cdot$10$^{34}$ cm$^{-2}$s$^{-1}$ and nominal length is set to 20 km. Two possible energy upgrades are foreseen, 500~\text{GeV} and 1 TeV center of mass energy, with the corresponding lengths of approximately 31 and 50 km, respectively. Also, dedicated scans of the WW$^\ast$ and $t\bar{t}$ processes at their respective resonant energies are foreseen, which will allow the determination of the W boson and t quark masses, with high precision. In the ILC H20 scenario \cite{ILCScenarios}, foreseen integrated luminosities at each energy stage, 250~\text{GeV}, 500~\text{GeV}, and 1 TeV, are 2, 4, and 8 ab$^{-1}$, respectively. 

This paper presents two studies of the statistical potential of ILC for the measurement of the branching fraction of the Higgs boson decay to WW$^\ast$, in the fully hadronic decay mode, using the Higgsstrahlung, $e^{+}e^{-} \rightarrow$ HZ, Higgs production channel. The studies are performed at two energy stages of ILC, 250~\text{GeV} and 500~\text{GeV} center of mass energy, with the integrated luminosity of 500 fb$^{-1}$ for each stage, using maximum beam polarization. 

\section{Detector design}

The ILC developed two detector concepts: the International Large Detector (ILD) \cite{ILDVersatile} and the Silicon Detector (SiD) \cite{SiD}. The main difference between the two detector options is in the tracking detectors. The ILD detector concept assumes gaseous tracking, a Time Projection Chamber (TPC), with silicon tracking for the outer and inner trackers. In contrast, the SiD detector concept developed an all-silicon tracking system. 
Both detectors are designed to follow the particle flow concept \cite{PFA1}\cite{PFA2} for the particle reconstruction to achieve resolution goals. The particle flow concept relies on the finely grained electromagnetic and hadronic calorimeters (ECAL and HCAL), as well as a highly pixilated vertex detector and sophisticated reconstruction algorithms.
The necessary resolutions of both detectors are motivated by the processes of interest \cite{ILDVersatile}. In particular, for the multi-jet final states, as studied in this paper, the jet-energy resolution is of primary importance. The latter is crucial for the separation of nearby jets from the W and Z bosons. Jet energy resolution is given by $\Delta E/E= 30\%/\sqrt(E)$\text{GeV}$^{-1}$ for the lower energy jets and around 3-4$\%$ for the jets with energy above 100~\text{GeV}. \\
The presented studies are performed using the GEANT4-based full simulation of the ILD detector.

\section{Higgs production at ILC}

\begin{figure}
    \centering
    \includegraphics[width=0.5\linewidth]{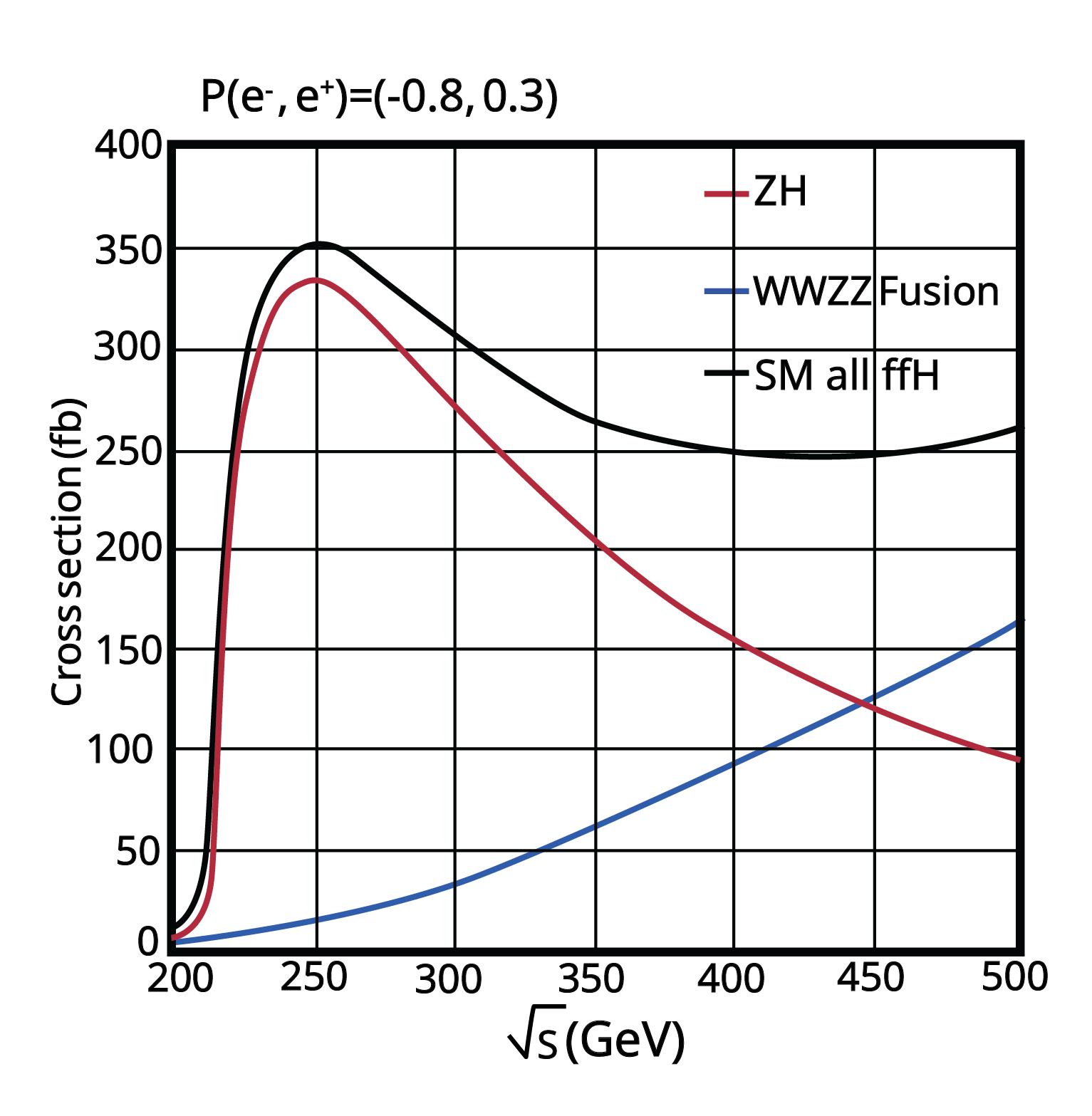}
    \caption{The cross section distributions of the Higgs boson production processes at the electron-positron colliders, as a function of the center of mass energy, for polarized beams $P(e^+e^-) = (+0.2,-0.8)$. The dominant Higgs production channels, the Higgsstrahlung and WW-fusion are depicted by the red and blue lines, respectively.}
    \label{fig:Cross-section}
\end{figure}

At the electron-positron colliders, the Higgs boson is predominantly produced via two concurrent processes, the Higgsstrahlung and the WW fusion. At lower energies, the leading Higgs production channel is the Higgsstrahlung, with the production cross section that peaks at around 250~\text{GeV} center of mass energy. This energy is chosen as the nominal energy for the first stage of ILC. The cross section of the sub-leading WW fusion process increases steadily with the center of mass energy and, at the foreseen second energy stage, 500~\text{GeV}, both Higgsstrahlung and WW fusion Higgs have comparable cross sections. At the last energy stage, 1 TeV, the WW-fusion dominates, with the sub-leading ZZ-fusion process. Figure \ref{fig:Cross-section} gives the cross sections of the Higgs boson production processes at electron-positron colliders, for the center of mass energies foreseen by the ILC.

\begin{table}[ht]
    \centering
    \begin{tabular}{lcccc}
\hline
             \multicolumn{2}{c}{Center of mass energy}  &  250~\text{GeV}    &  500~\text{GeV}   \\
Process                    & $P(e^+e^-)$  	& $\sigma[fb]$& $\sigma[fb]$\\
\hline
ee$\rightarrow$ HZ         & unpolarized                &  211 fb	  &   65 fb     \\
ee$\rightarrow$ Hvv        & unpolarized                &   21 fb	  &   72 fb     \\
ee$\rightarrow $HZ	       & (+0.3,-0.8)                &  318 fb	  &   96 fb     \\
ee$\rightarrow$ Hvv        & (+0.3,-0.8)                &   37 fb	  &  163 fb     \\
\hline
    \end{tabular}
    \caption{The cross-sections of the two leading Higgs production processes are given for unpolarized and fully polarized electron and positron beams.}
    \label{Table:signal}
\end{table}

To increase the effective Higgs production cross section at the ILC, polarization of both electron and positron beams is foreseen with the maximal polarization for electrons of 80$\%$ and for positrons 30$\%$. Table \ref{Table:signal} gives the cross sections of the Higgsstrahlung and WW fusion production processes, for beams without polarization and for maximally polarized beams, $P(e^+e^-)=(+0.3,-0.8)$. With nominal foreseen integrated luminosities in the ILC H20 scenario and maximally polarized beams, the production of more than 1.7$\times$10$^6$ Higgs bosons is expected at 250 and 500~\text{GeV} energy stages.

\section{Monte Carlo samples and analysis tools}

Signal and background samples are simulated using the Whizard 1.95 \cite{Whizard} event generator. The samples are generated using realistic beams that include luminosity spectrum and initial state radiation. The luminosity spectrum and beam-beam induced processes were simulated by GUINEA-PIG 1.4.4 \cite{GuineaPig}. The hadronization and fragmentation are simulated using PYTHIA 6.4 \cite{Pythia}. The hadronic background, resulting from the interaction of photons radiated from the incoming beams, is overlaid over each generated event before reconstruction. The detector response of the detector is simulated using GEANT4 with the full simulation of the ILD detector model, ILD$\_$o1$\_$v05, that is included in the ILCSOft \cite{ILCSoft}. Event reconstruction is done using the particle flow technique, implemented in the Pandora particle-flow algorithm (PFA) \cite{PFA1}\cite{PFA2}. The events are reconstructed using the MarlinReco package \cite{MarlinReco}.

The background is separated into several categories to study the influence of the event selection. The first two categories belong to the hadronic four-fermion final states that are produced through an intermediate state which contains two Z bosons or two W bosons. The third category includes the four-fermion final states that are produced either through a ZZ or WW intermediate state with the same quark content(“ZZ or WW”). The following two categories are specific semileptonic final states produced through ZZ or WW. The final two categories are the decays of the Higgs boson other than the signal and two-fermion processes. For the 500~\text{GeV} case, the additional category contains the six-fermion final states, which become relevant at this energy stage. Lists of signal and studied background processes, with their corresponding cross sections for both energy stages, 250 and 500~\text{GeV} center of mass energy, are given in Table \ref{Table:Background}, respectively.

\begin{table}
\centering 
\begin{tabular}{lrrrr}
\hline\noalign{\smallskip}
Center of mass energy [~\text{GeV}]                 	         &   \multicolumn{2}{c}{250~\text{GeV}}     &	\multicolumn{2}{c}{500~\text{GeV}}	     \\
    	                                                & $\sigma[fb]$    &	$\#$ evts      &   $\sigma[fb]$  & $\#$ evts\\
\hline
signal\\ 
$e^{+}e^{-}\rightarrow HZ, H\rightarrow WW^\ast \rightarrow q\bar{q}q\bar{q}, Z \rightarrow  q\bar{q}$ &  36.5& 18250& 11.3&5650 \\ 
\hline
background processes\\
\hline
$e^{+}e^{-}\rightarrow$ other Higgs decays         	     & 309.8 		   &      154900        &  103.4      &        51700\\ 
$e^{+}e^{-}\rightarrow $ 2f hadronic 	                 &19148.6		   &    64574300        &32470.5      &     16235250\\
$e^{+}e^{-}\rightarrow $ 4f WW hadronic 	             & 14874.3		   &     7437150        & 7680.7	  &      3840350\\
$e^{+}e^{-}\rightarrow $ 4f WW/ZZ hadronic 	             & 12383.3		   &     6191650        & 6400.1      &      3200050\\
$e^{+}e^{-}\rightarrow $ 4f ZZ hadronic 	             & 1402.0		   &	 701000         &  680.2   	  &       340100\\
$e^{+}e^{-}\rightarrow $ 4f WW semileptonic 	         & 18781.0 		   &     9390500        & 9521.4	  &      4760700\\
$e^{+}e^{-}\rightarrow $ 4f ZZ semileptonic 	         & 1422.1 		   &      711050        &  608.6	  &       304300\\
$e^{+}e^{-}\rightarrow $ 6f t$\bar{t}$ 		             & / 		       &        /           & 1338.6	  &       669300\\
\noalign{\smallskip}\hline\noalign{\smallskip}
\end{tabular}
\caption{List of considered processes with the corresponding cross sections at 250 and 500~\text{GeV} centre of mass energies, and number of events given for the assumed integrated luminosity of 500 fb$^{-1}$. }
\label{Table:Background}   
\end{table}

\section{Event reconstruction and preselection}

The studied process, ee$\rightarrow$HZ, H$\rightarrow$WW$^\ast$, contains three bosons in the final state, one on-shell and one off-shell W boson, that comprise the Higgs boson and the Z boson. The fully hadronic signal final state is investigated at both energy stages, thus the final state contains six jets. Due to the nature of the s-channel Higgsstrahlung Higgs production process, the produced jets are central. The jet opening depends on the center of mass energy. At lower energies, jets are widespread and potentially overlapping, which could lead to imperfect clustering of final-state particles into jets. At higher center of mass energies, the increased boost of jets leads to cleaner separation, resulting in a lower rate of particle-jet mismatching in jet clustering. 
However, at the higher energies, the presence of gamma-gamma to hadrons background influences the event reconstruction. 

In both analyses, event reconstruction of the six-jet final state was performed in two simultaneous procedures. First, the event is reconstructed by selecting the jet combination that provides the best reconstruction of the bosons (W, W*, and Z). Second, the jet openings are optimized for the jets in the event. 

The jet clustering is performed using the k$_t$ clustering algorithm as implemented in the FastJet v. 2.4.2 \cite{FastJet}. The reconstructed particles in the event are forced into six jets, each of the same cone opening (“jet radius”). To determine the jet openings for each energy stage, the jet radius is scanned from 0.8 to 1.5, in steps of 0.1. For each jet radius, the complete reconstruction of the event is performed: six jets of the event are paired to form three bosons of the event, the Z boson and real and virtual W bosons. The latter two are paired to reconstruct the Higgs boson. To obtain the best jet combination, a $\chi^2$ distribution is constructed, which takes into account only real bosons in the event, the Z, the W real and the Higgs boson:

\begin{equation}
\chi^2(W,Z,H) =\frac{(m_{ij}-m_W)^2}{\sigma^2_W} + \frac{(m_{kl}-m_Z)^2}{\sigma^2_Z} + \frac{(m_{ijmn}-m_H)^2}{\sigma^2_H}, \hspace{0.5cm} i,j,k,l,m,n=1,..6,
\label{Chi2}
\end{equation}

where m$_{ij}$ is the reconstructed mass of the Z boson candidate,  m$_{kl}$ is the reconstructed mass of the real W boson candidate, and m$_{klmn}$ is the invariant mass of the Higgs boson candidate, while m$_B$ and $\sigma_B$, (B = W, Z, H), are the masses and the expected mass resolutions of the corresponding bosons. The best jet combination is attained by the minimization of the constructed $\chi^2$ distribution.

After the event reconstruction for each jet opening, the fit of the reconstructed Z boson and the Higgs boson is performed. The fit is conducted in the vicinity of the expected boson masses (±5~\text{GeV} for the real W and the Z boson and ±10~\text{GeV} for the Higgs boson). The jet opening that provides minimal absolute error of the reconstructed Higgs boson and the Z mass to the nominal one is chosen.

\begin{figure}[ht!]
    \centering
    \begin{minipage}{0.5\textwidth}
        \centering
        \includegraphics[width=1.0\linewidth]{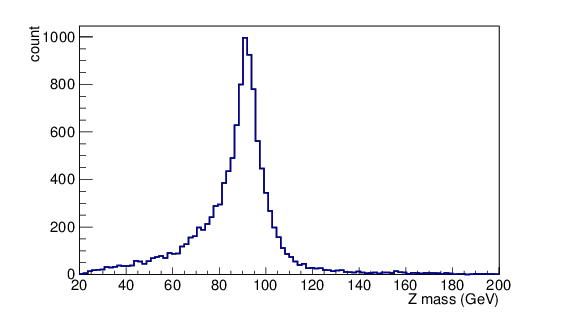} 
 
    \end{minipage}\hfill
    \begin{minipage}{0.5\textwidth}
        \centering
        \includegraphics[width=1.0\linewidth]{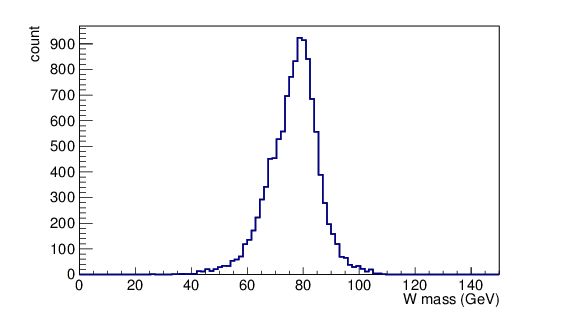} 
    \end{minipage}
           \caption{Distribution of the reconstructed invariant mass for the Z and W boson.}
\label{Fig:Masses}
\end{figure}

The best reconstruction of the invariant masses of the W real, the Z, and the Higgs boson for 250~\text{GeV} centre of mass energy is obtained for the jet radius of 1.5, while for 500~\text{GeV}, obtained jet radius is 1.2. The distributions of reconstructed invariant masses of the Z and the Higgs boson obtained by the best jet combination and jet opening are given in Figure \ref{Fig:Masses}.

\begin{figure}
    \centering
    \includegraphics[width=1.0\linewidth]{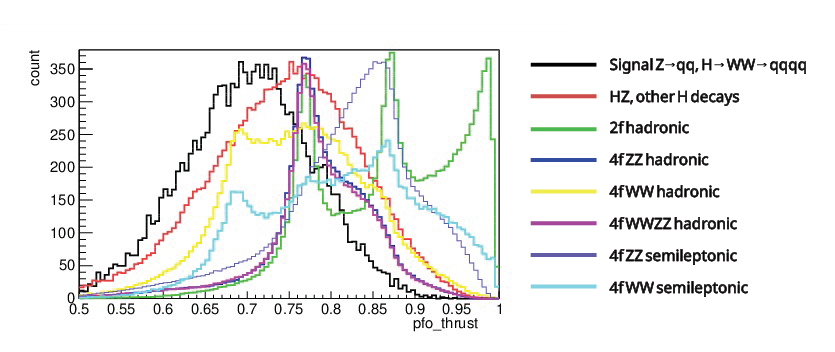}
    \caption{Reconstructed distributions of the event shape variable thrust, for signal (black) and backgrounds (colors) events after preselection (left) and final selection (right). Distributions have been rescaled to allow for direct shape comparison; the original event counts are not preserved.}
    \label{Fig:Thrust}
\end{figure}

In order to reduce semileptonic and high cross section two-fermion background, a set of preselection criteria is applied to the reconstructed signal and background events. The preselection uses several variables, of which one that describes the shape of an event in momentum space (thrust), as well as the jet transitions, y$_{ij}$, are the most effective for the background reduction. Jet transitions are k$_t$ algorithm values at which the algorithm makes the transition from the i to the j number of jets in the event. The distributions of the thrust variable, for signal and background, are given in the Figure \ref{Fig:Thrust}. Figure \ref{Fig:Thrust} illustrates the difference in distributions between the signal and background, particularly for di-jet final states. Clustering di-jet events into a six-jet topology produces a thrust distribution with three distinct peaks. These apparent peaks arise from the way the jet-finding algorithm subdivides the original two jets and from the effects of final-state radiation and gluon emission.

The first peak, near T$\approx$, corresponds to events in which all six reconstructed jets are aligned with the original jet directions. In this case, the algorithm subdivides each true jet into multiple smaller subjets within the same hemisphere, so the event largely retains its original two-jet topology.
The second peak, around T$\approx$0.87, arises from events in which the six reconstructed jets are somewhat misaligned, reflecting asymmetric energy distribution among the subjets.
The third peak, near T$\approx$0.87, results from final-state radiation and gluon emission, which perturb the original two-jet configuration. All distributions are scaled for shape comparison.

For the 250~\text{GeV} center of mass energy, the preselection criteria set include: event thrust $<$ 0.9, $-\log(y_{12}) < 2.2$, $-\log(y_{23}) < 3.0$, $-\log(y_{34}) < 3.5$, $-\log(y_{45}) < 4.0$, $-\log(y_{56}) < 4.0$, and $-\log(y_{67}) < 4.5$. Additional variables include the number of final state particles NPFO $<$ 70, the invariant mass of the $Z$ boson candidate $m_Z > 70$ ~\text{GeV}, the invariant mass of the Higgs boson candidate $m_H > 100$~\text{GeV}, the invariant mass of the real $W$ boson candidate $m_W > 60$~\text{GeV}, visible energy $E_{\mathrm{vis}} > 200$ ~\text{GeV}, and transverse momentum of a single jet $p_T > 20$~\text{GeV}.

With the rise of energy, jet separation improves, as does the reconstruction of the bosons in the event. For the second energy stage, 500~\text{GeV}, the topological features become more pronounced, making the event shapes between two jets and the six-jet signal more distinct. At this energy stage, the event thrust and reconstructed invariant mass of the $Z$ boson candidate are the leading discriminant variables. 

The preselection variable set contains the following criteria: event thrust $< 0.95$, the invariant mass of the $Z$ boson candidate $70~\mathrm{~\text{GeV}} < m_Z < 110~\mathrm{~\text{GeV}}$. Additional variables are the jet transitions: $-\log(y_{34}) < 3.7$, $-\log(y_{45}) < 4.3$, $-\log(y_{56}) < 4.5$, and the number of final state particles, NPFO $> 55$.

The number of signal and background events remaining after preselection for both energy stages is presented in Table \ref{Table:PreselectionEfficiencies}, along with the corresponding signal and background efficiencies.

\begin{table*}[t]
\begin{center}     
\begin{tabular}{lrrrrrr}
\hline\noalign{\smallskip}

Center of mass energy [~\text{GeV}]                    &   \multicolumn{3}{c}{250~\text{GeV}}   &	\multicolumn{3}{c}{500~\text{GeV}}	 	\\
    	                                     &  $\sigma[fb]$ &$\epsilon$[$\%$] &  $\#$ evts  &  $\sigma[fb]$    &	$\epsilon$[$\%$]	&	$\#$ evts    \\
\hline\noalign{\smallskip}
signal\\
$e^{+}e^{-}\rightarrow HZ\rightarrow WW \rightarrow qqqqqq$& 36.5	  &  	89.2		      &   16287     &	11.3           &79.7			&	  4515		\\ 
\hline\noalign{\smallskip}
background processes\\
\hline
$e^{+}e^{-}\rightarrow$ other Higgs decays 			        &309.8     &  	54.7		      &   84668		&     103.4          & 59.4			&    30710		\\ 
$e^{+}e^{-}\rightarrow $ 2f hadronic 			            &129148.6  &	 1.5	          &  950863		&   32470.5          &  3.4			&   551999		\\
$e^{+}e^{-}\rightarrow $ 4f WW hadronic 			 `      &14874.3   & 	33.5		      & 2491878		&    7680.7          & 29.1			&  1117542  	\\
$e^{+}e^{-}\rightarrow $ 4f WW/ZZ hadronic 		   	        &12383.3   &    33.8		      & 2092845		&	 6400.1          & 29.4			&   940815		\\
$e^{+}e^{-}\rightarrow $ 4f ZZ hadronic 			        & 	1402.0 &    42.9		      &  300880		&	 680.2           & 44.0			&   149644	 	\\
$e^{+}e^{-}\rightarrow $ 4f WW semileptonic 			    & 18781.0  &    0.005		      &     503		&	 9521.4          &0.0002		&	   952	 	\\
$e^{+}e^{-}\rightarrow $ 4f ZZ semileptonic 			    & 1422.1   &     0.5	          &    3557		&	 608.6           & 0.0005 		&	   137	 	\\
$e^{+}e^{-}\rightarrow $ 6f $t\bar{t}$ 			            & 	/ 	   &	 /                &   	/		&	1338.6           &  51.0		&   341317 	    \\

\noalign{\smallskip}\hline\noalign{\smallskip}
\end{tabular}
\caption{Preselection efficiencies and the number of events after the preselection, for $\sqrt{s}$ = 250~\text{GeV} and 500~\text{GeV} center of mass energies, assuming integrated luminosity of 500 fb$^{-1}$. } 
\label{Table:Preselection}
\end{center}
\end{table*}

\section{Machine learning based final selection}

The final discrimination between signal and background is performed using multivariate analysis, specifically the Boosted Decision Tree (BDT) method, as implemented in the TMVA package~\cite{TMVA}. 

Compared to leptonic and semileptonic decays, purely hadronic multi-jet final states lack highly efficient tagging variables such as the recoil mass distribution of the Higgs boson and the invariant mass of the $Z$ boson, which alone have high discriminating power \cite{ILCRecoil}. Thus, fully hadronic final states particularly benefit from statistical multivariate analysis methods, which can incorporate a higher number of sensitive variables to separate signal and background processes.

\subsection{Final selection at 250~\text{GeV}}

For the 250~\text{GeV} study, the BDT training is performed on purely hadronic backgrounds: non-$WW$ Higgs decays, $q\bar{q}$, and $q\bar{q}q\bar{q}$. The optimized set of training observables includes: the mass of the on-shell W boson; the mass of the Z boson; the mass of the Higgs boson; event shape variables such as thrust, aplanarity, oblateness, and sphericity; the number of reconstructed objects in the event; jet transition variables ($y_{12}$, $y_{23}$, $y_{34}$, $y_{45}$, $y_{56}$, and $y_{67}$); second highest $b$-tagging and $c$-tagging probabilities for the two jet hypotheses; transverse momentum of the highest $p_T$ jet; transverse momentum of jets comprising the Higgs boson; angle between the jets comprising the $Z$ boson; and the angle between the jets comprising the real $W$ boson. The variable set is optimized based on the criterion of minimal stable relative statistical uncertainty.

The most effective observables in the BDT training phase are the jet transition  $y_{45}$, $y_{56}$, $y_{67}$, and the thrust variable, which are connected to the spatial event topology that differentiates the six-jet final state from lower jet multiplicity final states. The effectiveness of the remaining variables in the set gradually decreases, with the last one being 10$\%$ efficiency. BDT selection efficiencies for signal and background are given in the Table \ref{BDT}. 

The BDT response is set to the cutoff value that maximizes the statistical significance:

\begin{equation}
S=\frac{N_S}{\sqrt{N_S+N_B}},
\label{formula:significance}
\end{equation}

  where N$_S$ and N$_B$ represent the number of selected signal and background events. Minimal relative uncertainty is obtained by maximizing significance $\Delta\sigma/\sigma=1/S$.\\
  
The overall efficiency of the signal, including preselection and final selection, is approximately 30$\%$, while the BDT efficiency is 35.7$\%$. The relative statistical uncertainty of the measurement $H \rightarrow WW \ast$ at 250~\text{GeV} center of mass energy is 4.1$\%$ using 500 fb$^{-1}$ of data. Considering ILC H20 scenario and integrated luminosity of 2 ab$^{-1}$ the relative statistical uncertainty improves to 2.0$\%$. 

\begin{table*}
\begin{center}     
\begin{tabular}{lrcrrrr}
\hline\noalign{\smallskip}

Center of mass energy [~\text{GeV}]                 	    &   \multicolumn{3}{c}{250~\text{GeV}}   &	\multicolumn{3}{c}{500~\text{GeV}}	 	\\
    	                                      		        &$\epsilon_{BDT}$[$\%$] &$\epsilon$[$\%$] &  $\#$ evts  & $\epsilon_{BDT}$[$\%$]   &	$\epsilon$[$\%$]		&	$\#$ evts    \\
\hline\noalign{\smallskip}
signal\\
$e^{+}e^{-}\rightarrow HZ\rightarrow WW \rightarrow qqqqqq$& 35.7	              &  	30.0		  &  5600       &   28.5    &22.7	&	1285  		\\ 
\hline\noalign{\smallskip}
background processes\\
\hline
$e^{+}e^{-}\rightarrow$other Higgs decays 			    &   7.5                &  	4.1	           &   6338		&   5.64   & 3.4          & 1733  \\  
$e^{+}e^{-}\rightarrow $2f hadronic 			        &   0.6                &	$<$10$^{-2}$   &   5410		&   0.34   & 0.010            & 1683  \\
$e^{+}e^{-}\rightarrow $4f WW hadronic 			 `      &   0.6                & 	0.2	           &   14961    &   0.055  & 0.017             & 616  \\
$e^{+}e^{-}\rightarrow $4f WW/ZZ hadronic 		   	    &   0.6                &    0.2	           &   13340    &	0.046  & 0.013             & 431  \\
$e^{+}e^{-}\rightarrow $4f ZZ hadronic 			        &   0.6                &  	1.0	           &    7178	&	0.151  & 0.066             & 226  \\
$e^{+}e^{-}\rightarrow $4f WW semileptonic 			    &$<$10$^{-5}$          &        /          &      / 	&	0.028  & $<$10$^{-4}$&  27  \\
$e^{+}e^{-}\rightarrow $4f ZZ semileptonic 			    &   1.2                &   $<$10$^{-2}$    &      49	&	$<$10$^{-3}$     & $<$10$^{-4}$ & /    \\
$e^{+}e^{-}\rightarrow $ 6f $t\bar{t}$ 			        & 	/ 	               &	    /        &   	/		&	0.266   & 0.13              & 907  \\
\noalign{\smallskip}\hline\noalign{\smallskip}
\end{tabular}
\caption{BDT and total selection efficiencies for signal and background processes at $\sqrt{s} = 250$ and $500~\mathrm{~\text{GeV}}$. The number of events after final selection is also given, assuming an integrated luminosity of $500~\mathrm{fb}^{-1}$.} \label{Table:PreselectionEfficiencies}
\label{BDT}
\end{center}
\end{table*}
  
\subsection{Final selection at 500~\text{GeV}} 
    
At 500~\text{GeV}, the final event selection is performed using all the relevant backgrounds: hadronic two and four-fermion hadronic background, other Higgs decays and $t\bar{t}$. BDT is trained with the following input observables: mass of the on-shell W boson; mass of the Z boson; mass of the Higgs boson; event shape variables: thrust, aplanarity, sphericity; number of reconstructed objects in an event; jet transition variables ($-\log y_{12}$, $-\log y_{23}$, $-\log y_{34}$, $-\log y_{45}$, $-\log y_{56}$ and $-\log y_{67}$); four highest b-tagging and c-tagging probabilities for the six jet hypothesis;  transverse momentum of jets that comprise the Higgs boson; visible energy; angle between the jets that comprise the Z boson;  angle between the jets that comprise real W boson; angle between the jets the reconstructed Higgs boson and the Z boson. The variable set is optimized in the same procedure as in the previous study. In addition to the variables that characterize event topology, one of the most effective variables at this energy is the invariant mass of the Higgs boson, which additionally addresses the separation of the six-jet final state of the signal and the $t\bar{t}$ background. 

\begin{figure} [b!]
    \centering
    \includegraphics[width=1.0\linewidth]{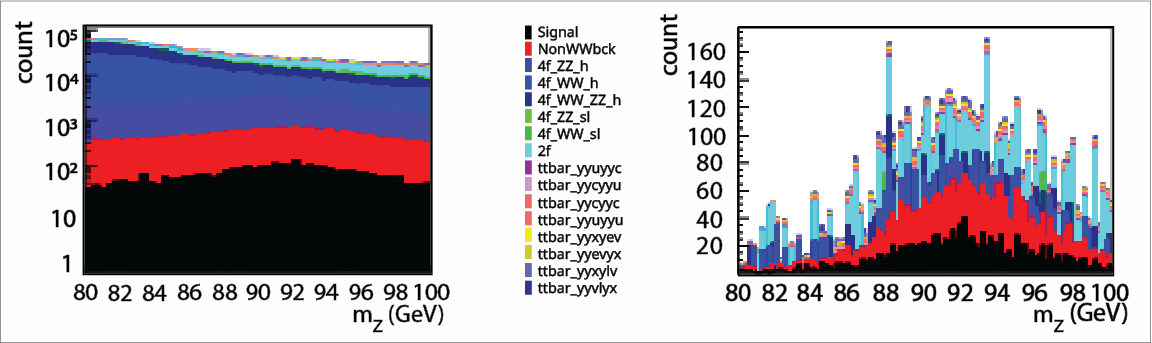}
    \caption{Distribution of the reconstructed invariant mass for the Z boson candidate for signal (black) and background (colors) after preselection (left) and final selection (right).}
 \label{Fig:Stackplot500}
\end{figure}

As in the 250~\text{GeV} case, the BDT response maximizes the statistical significance S. The obtained relative statistical uncertainty of the measurement is 6.5$\%$. The overall signal efficiency is 22$\%$, where the preselection signal efficiency is approximately 80$\%$, while the BDT signal efficiency is 30$\%$, with the background rejection efficiency greater than 99.9$\%$. The distributions of the invariant mass of the reconstructed Z boson for signal and background after the preselection and after the final selection are given in Figure \ref{Fig:Stackplot500}. The dominant contributions come from the two-fermion background and other decay channels of the Higgs boson. The number of signal and background events remaining after the final selection is given in Table \ref{BDT}, with the corresponding signal and background efficiency. The relative statistical uncertainty of the measurement $H \rightarrow WW^{\ast}$ at 500~\text{GeV} center of mass energy is 6.5$\%$ using 500 fb$^{-1}$ of data. Considering ILC H20 scenario and integrated luminosity of 4 ab$^{-1}$ the relative statistical uncertainty improves to 2.3$\%$.

\section{Conclusion}

  In this study, the statistical potential of the ILC for the measurement of the branching fraction of the Higgs boson to the pair of W bosons is evaluated based on a full simulation of the ILD detector for the first two energy stages of the ILC, $\sqrt{s}$ = 250~\text{GeV} and 500~\text{GeV} center of mass energy. Data sets are obtained using the realistic beams with the beam polarization of $P_{e^+e^-}=(+0.3,-0.8)$. In both studies, fully hadronic final states are considered. The obtained relative statistical uncertainties of the $\sigma$(HZ)x BR($H\rightarrow WW^{\ast}$) are 4.1$\%$ and 6.5$\%$ for the 250 and 500~\text{GeV}, respectively, assuming an integrated luminosity of 0.5 ab$^{-1}$ in both studies. Considering the ILC H20 scenario using nominal luminosities for the 250 and 500~\text{GeV} center of mass energy,  of 2 ab$^{-1}$ and 4 ab$^{-1}$, the obtained relative statistical uncertainties are 2.0  and 2.3$\%$ respectively. The final precision of the Higgs to W bosons coupling is determined in a global fit that combines the individual coupling measurements obtained at each energy stage of ILC. 
    
\acknowledgments

The author would like to express our gratitude to the LCC Generator Working Group and the ILD Software Working Group for supplying the simulation and reconstruction tools, as well as for generating the Monte Carlo samples used in this study. The author also acknowledges the support from the Ministry of Education, Science, and Technological Development of the Republic of Serbia.

\end{document}